\documentclass[noeprint,twocolumn,showpacs,amsmath,amssymb,sort&compress,floatfix,superscriptaddress,prl,aps]{revtex4-2}

\usepackage{graphicx}
\usepackage{dcolumn}
\usepackage[normalem]{ulem}
\usepackage{bm}
\usepackage[hypertexnames=false]{hyperref}
\usepackage[mathlines]{lineno}
\usepackage{mathtools}
\usepackage{amsmath}
\usepackage{float}
\usepackage[caption=false]{subfig}
\usepackage[english]{babel}
\usepackage{xcolor}
\usepackage{xparse}
\newcommand{\fig}{Fig.~}
\newcommand{\figs}{Figs.~}
\newcommand{\eq}{Eq.~}

\newcommand{\bl}[1]{{\color{black}#1}}

\newcommand{\ah}[1]{{\color{black}#1}}

\newcommand{\dm}[1]{{\color{black}#1}}
\newcommand{\dmN}[1]{{\color{black}#1}}
\newcommand{\mc}[1]{{\color{black}#1}}
\newcommand{\mcN}[1]{{\color{black}#1}}
\newcommand{\mcm}[1]{{\color{black}#1}}

\makeatletter
\def\maketitle{
\@author@finish
\title@column\titleblock@produce
\suppressfloats[t]}
\makeatother

\newcommand{\ucsb}{Department of Physics, University of California Santa Barbara, Santa Barbara, CA 93106, USA}
\newcommand{\ue}{SUPA, School of Physics and Astronomy, University of Edinburgh, Peter Guthrie Tait Road, Edinburgh EH9 3FD, United Kingdom}

\begin{document}

\title{Yield Stress and Compliance in Active Cell Monolayers}

\author{Austin Hopkins}
\affiliation{\ucsb}
\author{Michael Chiang}
\affiliation{\ue}
\author{Benjamin Loewe}
\affiliation{\ue}
\author{Davide Marenduzzo}
\affiliation{\ue}
\author{M. Cristina Marchetti}
\affiliation{\ucsb}

\date{\today}

\begin{abstract}
The rheology of biological tissue plays an important role in many processes, from organ formation to cancer invasion. Here, we use a multi-phase field model of motile cells to simulate active microrheology within a tissue monolayer. When unperturbed, the tissue exhibits a transition between a solid-like state and a \dm{fluid-like} state tuned by cell motility and deformability -- the ratio of the \mc{energetic} costs of steric cell-cell repulsion and \mc{cell surface tension.} 
\mcm{When perturbed,} solid tissues exhibit \dm{yield-stress behavior, with a threshold force for the onset of motion of a probe \mc{particle} that vanishes upon approaching the solid-to-liquid transition.} 
\mc{This} onset of motion is qualitatively different in the low and high deformability regimes. At high deformability, the tissue is amorphous when solid, it responds compliantly to deformations, and 
\mc{the probe transition to motion} is smooth. At low deformability, the monolayer is more ordered translationally \mc{and} stiffer, 
and the onset of motion appears discontinuous. 
\dm{Our results suggest that cellular or nanoparticle transport in different types of tissues can be fundamentally different, and point to ways in which it can be controlled.}  
\end{abstract}


\maketitle


The dynamics of cells in dense tissues is important for understanding many biological processes, including embryonic development \cite{Chuai2012}, cancer metastasis \cite{Haeger2014}, and wound healing \cite{Poujade2007}. \mcm{It underlies the}
 epithelial-mesenchymal transition  observed \textit{in vivo} \cite{Thiery2009,Thompson2005,Mitchel2020}, in which \dm{stationary} epithelial cells change to a more motile, mesenchymal phenotype. 
Experiments have also demonstrated a transition from glassy, \dm{or solid-like}, to liquid dynamics in epithelial monolayers both \textit{in vitro} \cite{Atia2018,Malinverno2017,Garcia2015,Park2015,Nnetu2012,Angelini2011} and \textit{in vivo} \cite{Atia2018,Mongera2018}.
Theoretical work on various models of dense tissues, including multi-phase field \cite{Loewe2020}, Voronoi \cite{Bi2016,Giavazzi2018}, vertex \cite{Bi2015,Barton2017,Li2018}, and cellular Potts models \cite{Chiang2016,Durand2019}, \mcm{has} shown that this melting transition can be driven by the interplay of \dm{cell surface} tension, cell motility, and active noise. 
\dm{An important question is whether the solid-to-liquid transition has an impact in tissue function in health and disease.}  Theory and experiments have begun to address this issue by exploring the rheological and mechanical properties of biological tissues, 
which have key consequences to their macroscopic biophysical behavior~\cite{Marmottant2009,Sandersius2011,sadeghipour2018shear,Mongera2018,Kim2021,tong2021linear,prakash2021motility,huang2021shear,hernandez2021geometric}.
\dm{A mechanistic and quantitative understanding of the
impact of cell \dm{surface} tension and cell motility on the rheology and transport properties of biophysical tissues is, however, still lacking}.

\mcm{To shed light on this  aspect, in this paper} we use active microrheology \cite{Squires2005}, \mcm{which measures the local response of the tissue to the drag of an embedded colloidal probe particle (Fig.~\ref{fig:vf_ex}),} to study the \mcm{viscoelastic} response of \mcm{a model tissue} monolayer.
Active microrheology has been used to study \mcm{the local material response in a} wide variety of \mcm{active and passive systems}, including colloidal suspensions~\cite{Habdas2004,Wilson2009,Puertas2014,Gruber2016,Gruber2020}, biological tissues \textit{in vivo}, \cite{DAngelo2019} \textit{in vitro}, and \textit{in silico} \cite{Sandersius2011}, and simulations of active disks \cite{Reichhardt2015,Burkholder2020,Knezevic2021}.
\mcm{While distinct from macrorheology that measures the material response on macroscopic scales, the two methods often yield qualitatively similar behavior when used to probe the rheology of complex fluids~\cite{Mohan2014}.}
Active microrheology is an especially promising technique in biological systems where it may be less destructive than common macrorheology experiments. 

\begin{figure}
    \centering
    \includegraphics[width=0.9\linewidth]{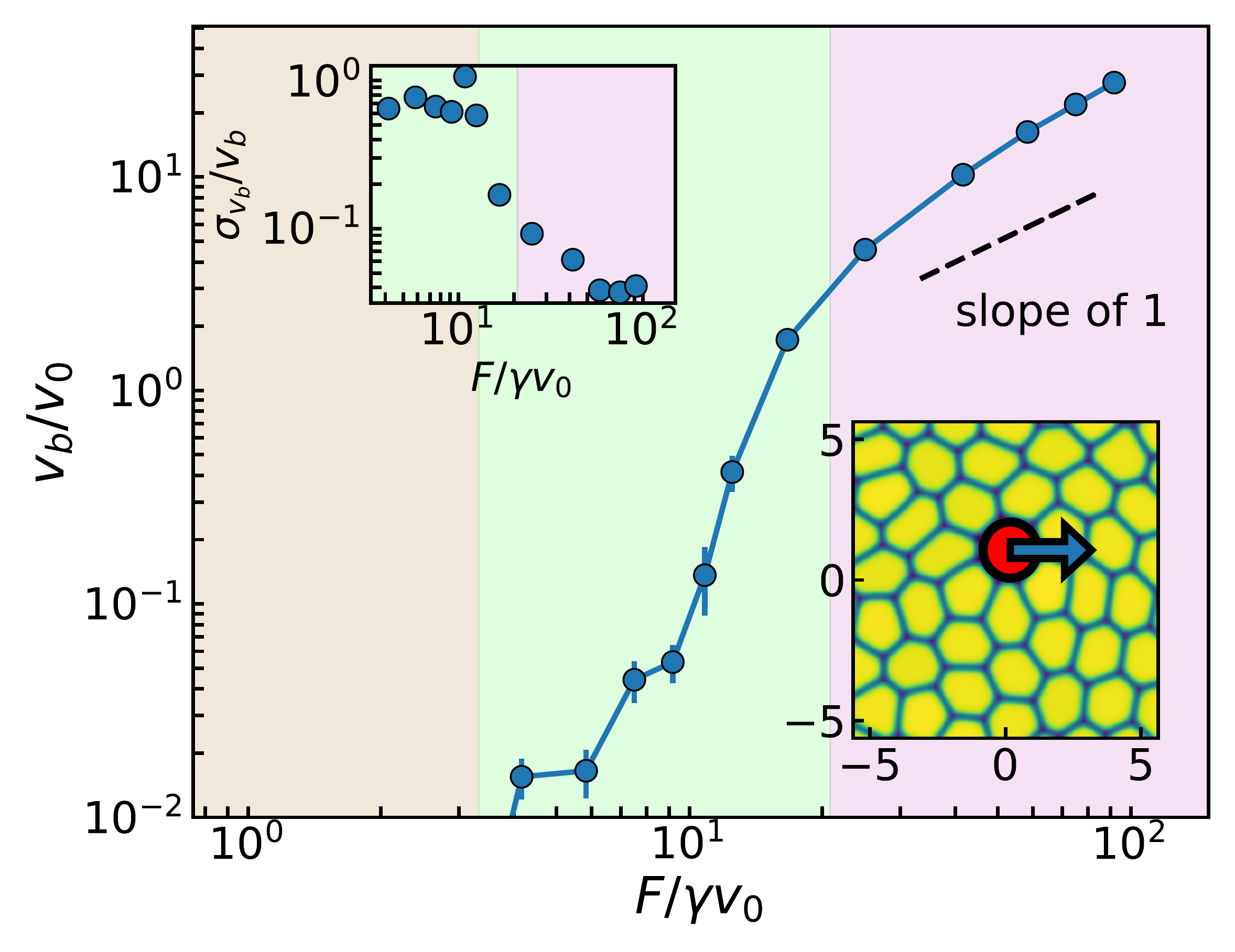}
    \caption{A typical velocity-force curve for a hard probe particle in a glassy tissue ($d = 3$, $\text{Pe} = 0.1$) showing three dynamical regimes [from left to right: caging (brown), stick-slip (green), and moving freely (purple)]. Upper inset: the relative strength of the \mc{velocity} fluctuations decreases above the critical force. Lower inset: \mc{a} snapshot of the simulation setup.}
    \label{fig:vf_ex}
\end{figure}

\mcm{Using a multi-phase field model \bl{\cite{Nonomura2012,Palmieri2015,Mueller2019,Peyret2019,Loewe2020,Zhang2020,Wenzel2021,Zhang2021,Monfared2021}} of cells as deformable active Brownian particles (ABPs) on a substrate, we show that \dm{ solid-like} tissues exhibit a finite threshold for the onset of motion of an \dm{embedded} colloidal probe pulled at a constant force. This threshold force can be interpreted as a measure of the yield stress of the tissue.  \dm{One of our main findings is that the} nature of the yielding transition depends qualitatively on cell deformability, \dm{or surface tension}. 
Soft tissues of highly deformable cells are compliant and adapt to the deformation, resulting in a smooth onset of motion and small yield stress. In  contrast, in tissues composed of rigid cells, the probe induces both deformations and local translations of cells in its immediate neighborhood, and the onset of motion appears discontinuous, with large yield stress. \dm{The difference is also visible in the spatial patterns of cell deformations and stress transmission in the neighborhood of the probe, which are fundamentally distinct in the two cases. The reason underlying the different behavior is that the solid-like phase is amorphous for high deformability, and close to crystalline for small ones. As we discuss at the end of our work, these results have implications for the transport of cells and nanoparticles in different types of tissues.}}

\paragraph{\mcm{Model.}} \mcm{We describe \dm{a tissue monolayer} as a collection of $N$ cells modeled as deformable ABPs, each} identified by a phase field $\phi_i(\bm{r})$, with $i=1,\dots,N$. The free energy of the system is given by
\begin{equation}
\begin{split}
    \mathcal{F} &  = \sum_{i=1}^N \bigg[ \int d^2\bm{r} \left(\frac{\alpha}{4}\phi_i^2(\phi_i - \phi_0)^2 + \frac{K}{2} (\bm\nabla \phi_i)^2 \right) \\ +
     & \lambda \left( 1 - \int d^2\bm{r}\, \frac{\phi_i^2}{\pi R^2 \phi_0^2} \right)^2 + \epsilon \sum_{i<j=1}^N \int d^2\bm{r}\, \phi_i^2 \phi_j^2 \bigg].
\end{split}
\label{eq:fe}
\end{equation}
%
The first term sets $\phi_0=2$ and $0$ as the preferred values of the interior and exterior of each cell, respectively. The second term penalizes gradients in the phase with a stiffness  $K$.
These two terms determine the interfacial thickness $\xi=\sqrt{2K/\alpha}$ and the \dm{cell surface} tension $\sigma=\sqrt{8K\alpha/9}$. 
The third term is a soft constraint on the area of the cell, setting its preferred \mcm{area} to that of a circle of radius $R=12$. 
Finally, the fourth term models steric repulsion \mc{($\epsilon = 0.1$)} by  penalizing cell overlap.

Cell dynamics is overdamped due to friction with the substrate and is governed by the equation
\begin{equation}
    \frac{\partial \phi_i}{\partial t} + \bm{v}_i \cdot\bm\nabla \phi_i = -\frac{1}{\gamma} \frac{\delta \mathcal{F}}{\delta \phi_i}\;,
\label{eq:odd}
\end{equation}
where \mc{$\gamma = 10$} is \mcm{an inverse mobility. Cell motility enters}  through advection by the cell's \mcm{self-propulsion velocity $\bm{v}_i = v_0 (\cos\theta_i,\sin\theta_i)$.}
As in models of rigid ABPs, \mc{we assume all cells move at the same speed $v_0$}, while their direction of motion $\theta_i$ evolves independently, performing a random walk with rotational diffusion coefficient $D_r$, \mc{i.e.,} 
%
%
%
\begin{equation}
    d\theta_i(t) = \sqrt{2 D_r}\,dW_i(t),
\label{eq:theta}
\end{equation}
where $W_i(t)$ is a Wiener \mc{process} \mc{and $D_r = 10^{-4}$}. 
We quantify cellular activity through the P\'eclet number $\text{Pe} = v_0/(R D_r)$, which is the ratio between the cells' persistence length $v_0/D_r$ and their size $R$.

When cells interact, they may overlap or deform their shape. 
We quantify \mcm{cell} deformability \mcm{in terms of the dimensionless parameter} $d=\epsilon \xi R/(\sigma R)=3\epsilon/(2\alpha)$, which measures the ratio between the energy scales of overlap and \dm{surface tension}. 
In our simulations, we vary $d$ from $0.3$ to $6.0$ by changing the tension $\sigma$ for a fixed interface width \mc{$\xi = 2$}.
\mc{The compressibility $\chi=\lambda/(\epsilon \xi R)$  characterizes the competition between area changes and overlap and is fixed to $\chi=125/3$, which yields polydisperse systems. Further details of the implementation of the model were discussed previously in~\cite{Loewe2020}.}

To simulate \mcm{active} microrheology experiments, we \mcm{embed in the tissue} a probe particle \mcm{that is also described by a phase field and \mc{subjected} to} the same free energy [\eq\eqref{eq:fe}] as the cells, but with a small deformability ($d=0.015$) and strong area constraint ($\chi = 2500$), 
so that it is effectively rigid and \mcm{remains circular at all times}. \mcm{The probe is endowed with tunable self-propulsion velocity directed along the $x$ axis}, effectively reproducing a constant force in the $+x$ direction, as shown in the \mc{lower} inset of \fig\ref{fig:vf_ex}.
We measure the probe's velocity as $v_b(t)=\frac{x_b(t+\Delta t) - x_b(t)}{\Delta t}$, \mcm{where $x_b(t)$ is the instantaneous $x$ position of the probe,} over intervals of $\Delta t = 0.01 D_r$ for a range of values of $d$ and $\text{Pe}$.

We characterize the \dm{solid-} or \mc{liquid-like} state of the monolayer by examining the long-time behavior of the mean squared displacement (MSD), with $\text{MSD}(t)\sim t^\alpha$. In the \dm{solid-like} state the cells behave subdiffusively \mc{($\alpha < 1$),} 
whereas in the \mc{liquid-like} state \mc{they} move diffusively ($\alpha = 1$; \mc{see also \figs \ref{fig:msds_v0} and \ref{fig:msds_d}).}

\begin{figure*}[!th]
    \centering
    \includegraphics[width=\linewidth]{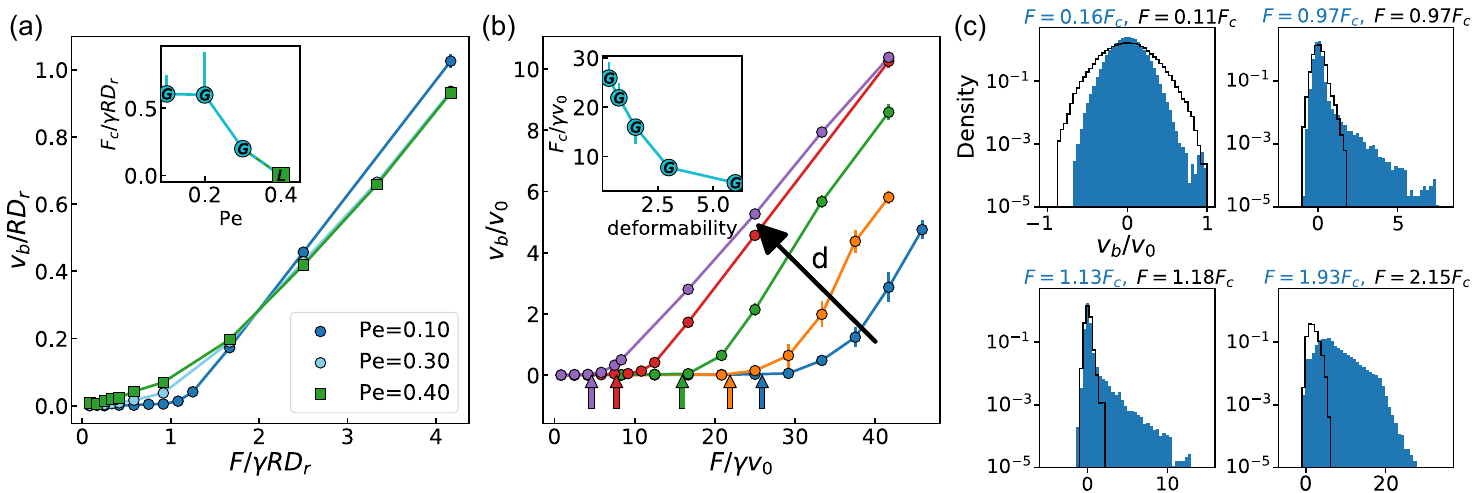}
    \caption{Probe velocity $v_b$ as a function of the applied force $F$ from varying cell motility and deformability. (a) Velocity-force curves at high deformability ($d=3$) for various  $\text{Pe}$. The onset of motion becomes increasingly rounded with higher $\text{Pe}$. Inset: The threshold force $F_c$ versus $\text{Pe}$. \mcm{Circles labeled $G$ denote a glassy state with a subdiffusive MSD, whereas squares labeled $L$} mark a liquid system with a diffusive MSD. (b) Velocity-force curves at $\text{Pe} = 0.1$ for various deformability values ($d = 0.3, 0.75, 1.5, 3.0$, and $6.0$). Inset: Threshold force versus deformability. All points correspond to the tissue being in a glassy state. (c) Histograms of $v_b$ for $d=0.3$ (solid blue) and $d=3.0$ (black outline). At low deformability the distributions are much broader and exhibit fat tails at high velocity for values of $F$ near the critical force. The blue text indicates the force for $d=0.3$, whereas the black text is for $d=3.0$.}
    \label{fig:fc_v0_and_d}
\end{figure*}

\paragraph{\mcm{Threshold force and yield stress.}} \mcm{\dm{In our solid-like tissue}, a finite threshold force $F_c$ is required for the probe to move at} a 
\mc{non-zero} velocity (\fig \ref{fig:vf_ex}). In colloidal suspensions, the threshold force probed by microrheology can be related quantitatively to the macroscopic yield stress \cite{Mohan2014}. \mcm{While the form of such a relation has not been established for the case of deformable particles, where macroscopic rheological measurements or simulations are not yet available, this suggests that the behavior of $F_c$ should be at least qualitatively similar to that of the \dm{tissue} yield stress. 

A typical \mc{velocity-force} curve for our rheological probe is shown in \fig\ref{fig:vf_ex}. We identify three dynamical regimes:}
\ah{At low forces, the probe rattles within its cage but is unable to escape its neighbors (light brown region).
At greater forces (green region), the probe is able to deform its neighbors strongly enough to escape its cage, but  can be temporarily trapped in new cages, resulting in stick-slip motion.
As the force is further increased, it no longer spends any time caged -- its instantaneous velocity becomes finite at all times, and the velocity-force curve eventually turns almost linear.
We identify the threshold force $F_c$ as the force at which the average steady-state displacement of the bead's position $\Delta x_b^{\textrm{late}}$ is at least $R$. \mcm{The choice of the cutoff, or a definition based on the late time exponent with which displacement grows with time, does not affect the qualitative behavior of $F_c$ as a function of $\text{Pe}$ and deformability [see Supplemental Materials (SM)].}
Below $F_c$ the probe is either completely caged or engages in very rare stick-slip motion.
Just above $F_c$, it instead moves substantially through the tissue, in frequent stick-slip motion or steady motion.}
\mcm{The strength of velocity fluctuations relative to the mean} decreases for $F>F_c$ (\mc{see the upper} inset of \fig\ref{fig:vf_ex}), which is further evidence for a dynamical transition \dm{associated with yielding}.


To examine the effect of activity, we vary \mc{$\text{Pe}$} 
at fixed deformability. 
The resulting \mc{velocity-force} curves are shown in \fig\ref{fig:fc_v0_and_d}(a) for highly deformable cells (\mc{$d = 3$}). \mcm{Increasing activity leads to a smoother transition at the onset of motion.} 
This is qualitatively similar to the thermal rounding observed in depinning phenomena \cite{Middleton1992}, \mcm{although in our system the \mc{probe's} velocity is zero over a finite range of applied forces for small $\text{Pe}$, indicating that the threshold force \mc{$F_c$} is non-zero even in the presence of noise.}
\mc{Specifically, $F_c$ and the associated yield stress decrease with $\text{Pe}$,}
but remain finite within the precision of our simulations until $\text{Pe} = 0.4$ \mc{[see the inset of \fig\ref{fig:fc_v0_and_d}(a)]}. \mcm{This is also the point \mc{at which} the tissue melts, as evidenced by measurements of MSD (\fig\ref{fig:msds_v0}),} suggesting that the existence of a finite yield stress can also be used to characterize the rheological state of the tissue.


\mcm{To examine} the effect of deformability, we fix $\text{Pe} = 0.1$ and vary $d$ [\fig\ref{fig:fc_v0_and_d}(b)].
For this value of $\text{Pe}$, the system remains \dm{solid-like} for all 
\mc{values of $d$} considered \mc{(\fig\ref{fig:msds_d})}. \mcm{\figs\ref{fig:fc_v0_and_d}(b) and \ref{fig:fc_d2} show that the onset of motion \dm{changes qualitatively depending on cell surface tension (or deformability)}. Tissues composed of cells with low 
tension that can be easily deformed are highly compliant and can adapt to the deformation induced by the probe simply through cell-shape changes. This yields a smooth, continuous onset of motion and low yield stress. In contrast, rigid cells that resist deformation result in a sharp, almost discontinuous onset of motion and large yield stress, as in this case the probe needs to push aside its neighbors to start moving. The decrease of threshold force with increasing deformability can be understood by assuming that $F_c\sim k_{\text{eff}}$, where $k_{\text{eff}}$ is  the effective spring constant felt by a caged particle. A calculation shows that $k_{\text{eff}}\sim 1/d$ (see SM), in line with the measurements shown in the inset of \fig\ref{fig:fc_v0_and_d}(b). \dmN{The qualitative difference in behaviour at low and high deformability becomes apparent when plotting the histograms of the instantaneous velocities of the probe, which are much broader and exhibit fatter tails for low $d$, especially in the vicinity of the threshold force [\fig\ref{fig:fc_v0_and_d}(c)].}} 

\begin{figure}
    \centering
    \includegraphics[width=\linewidth]{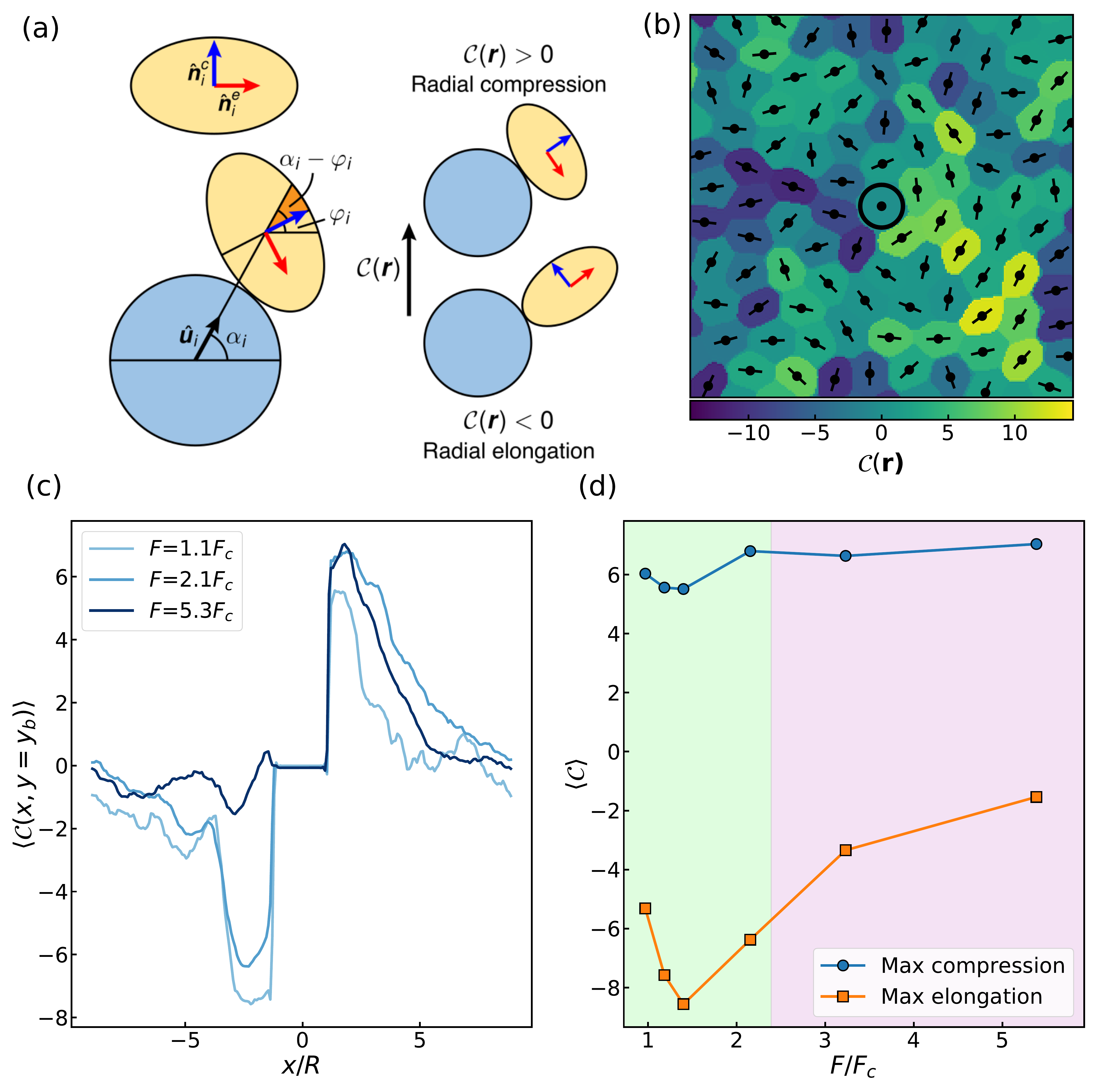}
    \caption{\mc{Quantifying the deformation of cells surrounding the probe. (a) A diagram explaining the radial compression field $\mathcal{C}(\bm{r})$, which is computed based on the alignment of the cell's compression axis $\hat{\bm{n}}_i^c$ with the unit vector $\hat{\bm{u}}_i$ pointing from the center of the probe (blue) to the center of mass of the cell (yellow).} 
    (b) Example of $\mathcal{C}(\bm{r})$ for $F=0.93F_c$ with the \mc{axis of elongation $\hat{\bm{n}}_i^e$ shown (perpendicular to $\hat{\bm{n}}_i^c$).} 
    (c) Profile of the compression field along the center of the probe in the $x$ direction, i.e., $\langle\mathcal{C}(x,y=y_b)\rangle$. Darker \mc{shades of} blue correspond to greater forces. (d) The minima and maxima of $\langle\mathcal{C}(x,y=y_b)\rangle$ as a function of the applied force. \mc{$d = 3$ and $\text{Pe} = 0.1$ for all plots in this figure.}}
    \label{fig:def_f}
\end{figure}

\paragraph{\mcm{Deformations and stress patterns~\dm{close to the probe.}}} \dm{As the probe is dragged around the tissue, it can substantially deform cells nearby (Figs.~\ref{fig:def_f} and~\ref{fig:def_f2}). To quantify} the extent of such deformations, 
\mc{we consider the traceless deformation tensor $\bm{S}_i$ \cite{Mueller2019}, with components 
\begin{align}
  S_{i,{\alpha\beta}} = -\int d^2\boldsymbol{r} \left[(\partial_{\alpha}\phi_i)(\partial_{\beta}\phi_i)-\frac{\delta_{\alpha\beta}}{2}(\partial_{\gamma}\phi_i)^2\right].
\end{align}
This tensor has} eigenvectors $\hat{\bm{n}}^e_i$ and $\hat{\bm{n}}^c_i$ with eigenvalues $s_i$ and $-s_i$, respectively, where $s_i=\sqrt{S_{i,xx}^2 + S_{i,xy}^2}$ gives the magnitude of the deformation. 
$\hat{\bm{n}}^e_i = \left(\sin\varphi_i,-\cos\varphi_i\right)$ points along the axis of greatest elongation, whereas $\hat{\bm{n}}^c_i = \left(\cos\varphi_i,\sin\varphi_i\right)$ is along the axis of greatest compression, \mc{and $\varphi_i$ is the angle between $\hat{\bm{n}}^c_i$ and the $x$ axis [\fig\ref{fig:def_f}(a)].} 
\mc{Using this tensor, we measure the degree of deformation of each cell in the direction from the probe by defining the radial compression field 
\begin{equation}
    \mathcal{C}(\bm{r}) = \sum_i s_i\,H(\phi_i(\bm{r})-\phi_0/2) \,\cos\left(2(\alpha_i-\varphi_i)\right),
\end{equation}
where $H$ is the Heaviside function and the field is smoothed to interpolate values on cell boundaries.} 
\mc{Here, $\alpha_i$ is defined by the unit vector $\hat{\bm{u}}_i = \left(\cos\alpha_i,\sin\alpha_i\right)$ 
that points from the center of the probe to the center of mass of cell $i$}. 
\dm{Positive or negative values of $\mathcal{C}(\bm{r})$, respectively, signify local compression or elongation \mc{along $\bm{r}$}.} 
\dm{\fig \ref{fig:def_f}(b) shows a snapshot of $\mathcal{C}(\bm{r})$,}
\mc{indicating} 
a buildup of compression along the $x$ direction in front of the probe and a wake of elongation behind.

\dm{To estimate the lengthscale over which the tissue is deformed, we \mc{average the compression field over time and different initial configurations. 
\fig\ref{fig:def_f}(c) shows a slice of 
this field}, $\langle\mathcal{C}(x,y=y_b)\rangle$, through the center of the probe along the direction of the pulling force.}
\mcm{Compression and elongation are largest at about} one cell length from the probe, independent of the force. 
\ah{The magnitude of deformation decays over a scale of order $2R$ for all forces.}
The compression remains roughly constant beyond the critical force; the probe induces enough of a compression to escape its cage.
The magnitude of elongation behind the probe first increases, then decreases with increasing applied force \bl{[\fig\ref{fig:def_f}(d)].}
\mcm{This nonmonotonic behavior arises because elongation requires that 
the probe to create free space where the cells behind it can expand.}
When the probe is undergoing stick-slip motion, \mcm{the large instantaneous forces create large empty regions behind \mc{the probe} 
in which nearby cells can elongate. Instead, when the probe moves freely, the cells behind it follow the motion smoothly and their elongation decreases.}

\begin{figure}
    \centering
    \includegraphics{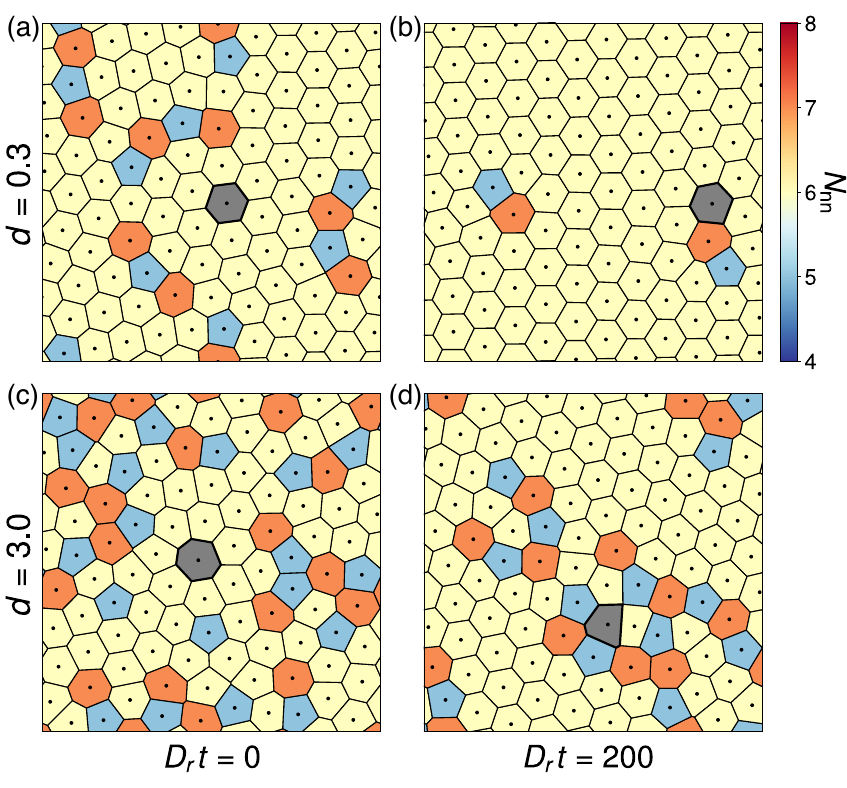}
    \caption{Voronoi tessellation of the system showing the decay in the number of disclinations over time at different deformability $d$. (a), (b) The number of \mc{nearest} neighbors \mc{$N_{\text{nn}}$} of each cell at time (a) $D_rt = 0$ and (b) $200$, respectively, for $d = 0.3$. (c), (d) \mc{$N_{\text{nn}}$} of each cell at the same two time points, but for $d = 3$. In all cases, a constant force of $F/(\gamma v_0) = 25$ is applied to the probe particle (colored in gray) in the $+x$ direction. Note that for $d = 3$, the probe has traveled across the vertical periodic boundary before the snapshot in (d).}
    \label{fig:defects}
\end{figure}

\mcm{The sharpness of the onset of motion is directly correlated with the structure of the \mc{solid-like} state, as evident from \fig\ref{fig:defects} that shows the defects obtained by Voronoi tessellation \mc{of the monolayer}. For high cell surface tension, corresponding to rigid cells, \mc{this} 
state~\dm{forms a regular hexagonal lattice, with few disclinations [\fig\ref{fig:defects}(a)]: 
the onset of motion appears discontinuous, and the yield stress is large.}}
For low cell surface tension, corresponding to deformable cells, the 
state is amorphous, with a large number of disclinations \mc{[\fig\ref{fig:defects}(b)]}. \mcm{In this case, the tissue is compliant and easily reconfigurable in response to perturbations, resulting in a smooth onset of motion and small yield stress.} In general, the number of disclinations also depends on initial conditions. \mcm{At all values of deformability, the motion of the probe tends to heal} the system, reducing the number of defects \mc{[\figs\ref{fig:defects}(c) and (d)]}.
\mcm{At high pulling forces the free motion of the probe acts as noise in the system, which promotes both annihilation and creation of defect pairs,} resulting in fluctuations in the number of disclinations over time \mc{(\fig\ref{fig:defects_time})}. 
\mcm{The behavior is qualitatively similar to that of two-dimensional extended system pinned by quenched disorder, where a uniform external drive first depins the system, setting it into motion, and then heals it at large applied forces, restoring partial translational order~\cite{giamarchi1994elastic,balents1998nonequilibrium}.   } 

\dm{To conclude, we have performed simulations of active microrheology in model cell monolayers, and shown that this method} \mcm{provides a useful tool for quantifying the emerging material properties of the system.} 
\dm{We found that the monolayer behaves as a yield-stress material, with a depinning-like yielding transition at a finite value of the pulling force.} \mcm{We also demonstrated that there is a fundamental relationship between the nature of the yielding transition and microscopic cell properties such as surface tension and motility. \dm{Monolayers formed by rigid cells are crystalline at rest, yield at a larger force, and the yielding transition is first-order-like. Monolayers formed by softer cells are amorphous, more compliant and yield at a smaller force, with a much smoother underlying transition. As the surface tension of cancerous cells is thought to be substantially larger than than of healthy tissues, our results suggest that transport and dynamics inside tissues may change dramatically in disease, and it would be of interest to test this prediction experimentally.}} \mcm{Our results are also relevant to recent dynamical measurements in \emph{Drosophila} embryos based on embedding a probe particle inside an individual cell~\cite{DAngelo2019}, or ferroelectric droplets in between neighboring cells~\cite{Mongera2018}, which have begun to show that local probes can provide quantitative information on local tissue rheology \emph{in vivo}.} 


\begin{acknowledgments}
The work by A.H. and M.C.M. was supported by the National Science Foundation Grant No.~DMR-1720256 (iSuperSeed)  with additional support from DMR-2041459. \bl{This research has received funding (B.L.) from the European Research Council under the European Union’s Horizon 2020 research and innovation programme (grant agreement No. 851196).}
\end{acknowledgments}

\bibliography{microrheology}

\pagebreak
\cleardoublepage 


\title{Yield Stress and Compliance in Active Cell Monolayers: Supplemental Materials}

\maketitle


\setcounter{equation}{0}
\renewcommand{\theequation}{S\arabic{equation}}

\setcounter{figure}{0}
\renewcommand{\thefigure}{S\arabic{figure}}

\section*{Additional Simulation Details}
Unless otherwise specified, we initialize the system with random non-overlapping positions for $N=100$ cells in a $215$ by $215$ simulation box, corresponding to a density $\rho = \frac{\text{total target area}}{\text{simulation box}} \approx 0.98$, i.e., a tissue that is almost confluent. The initial cells begin as circles with a radius smaller than their preferred radius. We first do a passive run to allow the cells to relax and grow, then turn on the motility of the tissue cells, and then turn on the force applied to the bead. We use a time step $\Delta t = 0.1$ to ensure stability of the bead. For all systems, we average over 10 runs of the bead in different initial configurations, unless $d=0.2$ where we average over 20 runs.
We run additional short runs to collect better statistics for the radial compression. In determining the mean squared displacements below, we use a larger time step $\Delta t = 0.5$ and average over 3 runs.

\section*{Mean Squared Displacement}
\begin{figure*}[!th]
    \centering
    \includegraphics[width=0.49\linewidth]{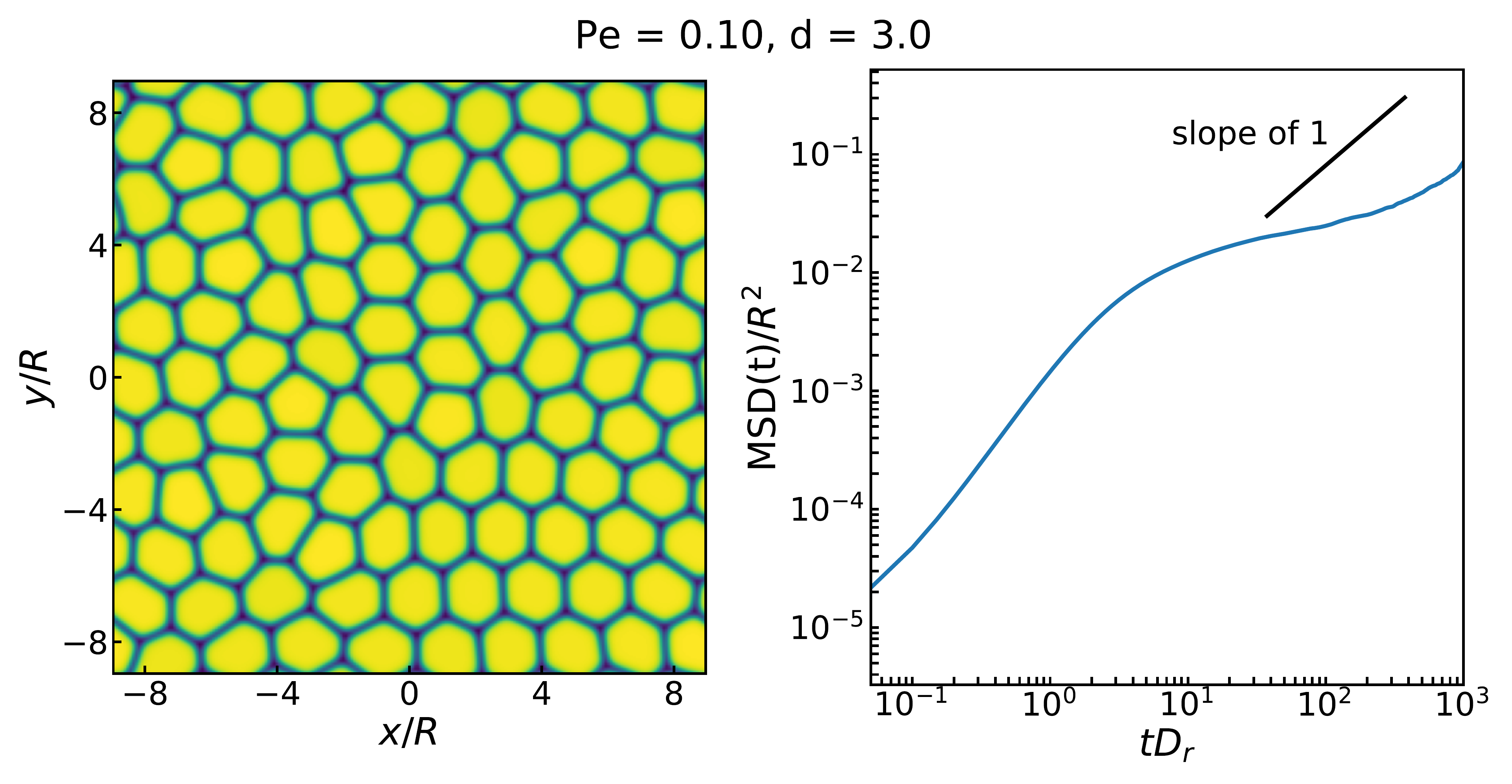}
    \includegraphics[width=0.49\linewidth]{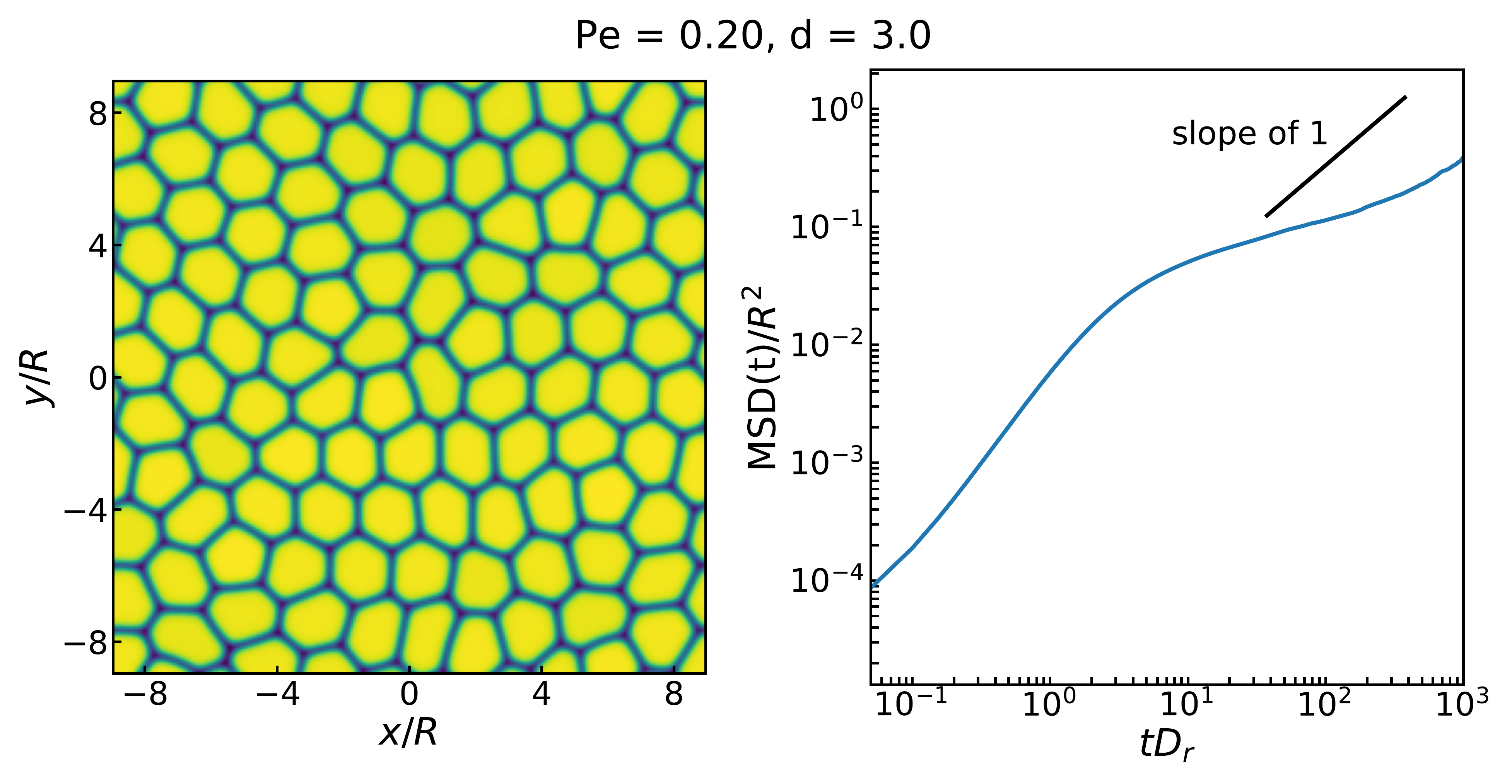}
    \includegraphics[width=0.49\linewidth]{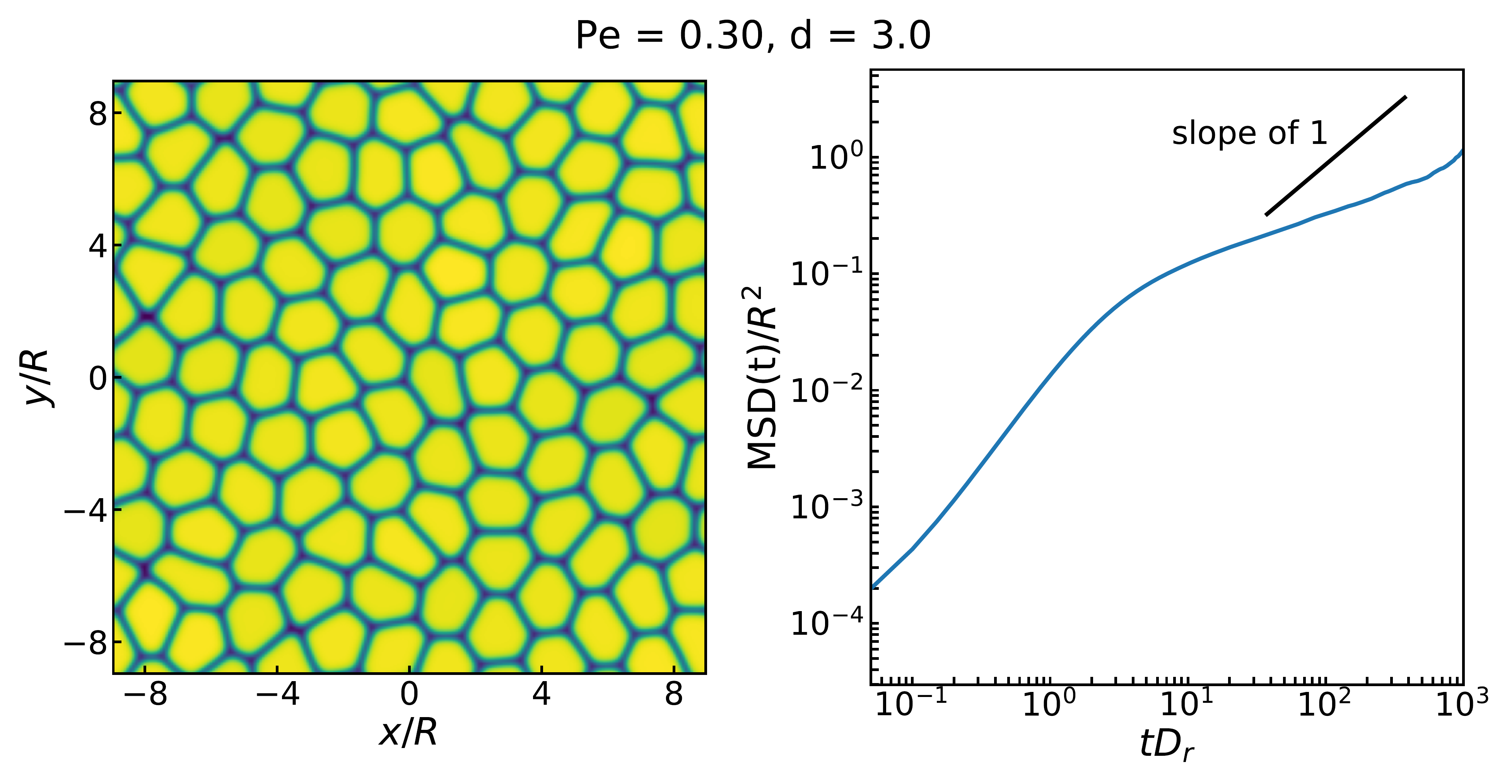}
    \includegraphics[width=0.49\linewidth]{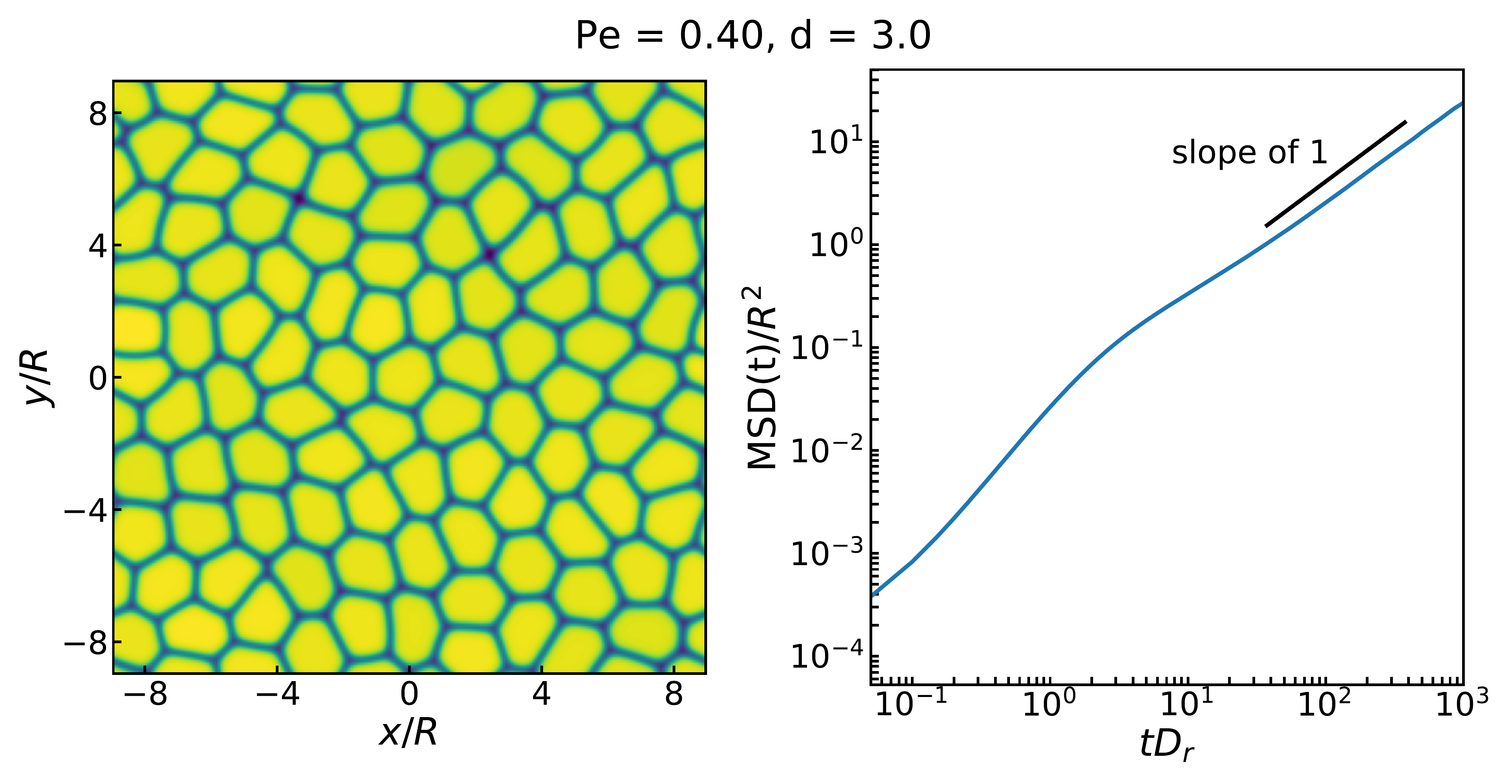}
    \caption{Snapshots and mean squared displacements \mc{(MSDs) for systems with fixed $d = 3.0$ and $\text{Pe}$ varied from $0.1$ to $0.4$, as considered in \fig\ref{fig:fc_v0_and_d}(a). Here, MSDs are plotted in log-log scale and show that as ${\text{Pe}}$ increases, the slope of the curve (i.e., the long-time exponent $\alpha$) becomes closer to $1$, indicating that the monolayer changes from a solid-like to a fluid-like state.}}
    \label{fig:msds_v0}
\end{figure*}

\begin{figure*}[!ht]
    \centering
    \includegraphics[width=0.49\linewidth]{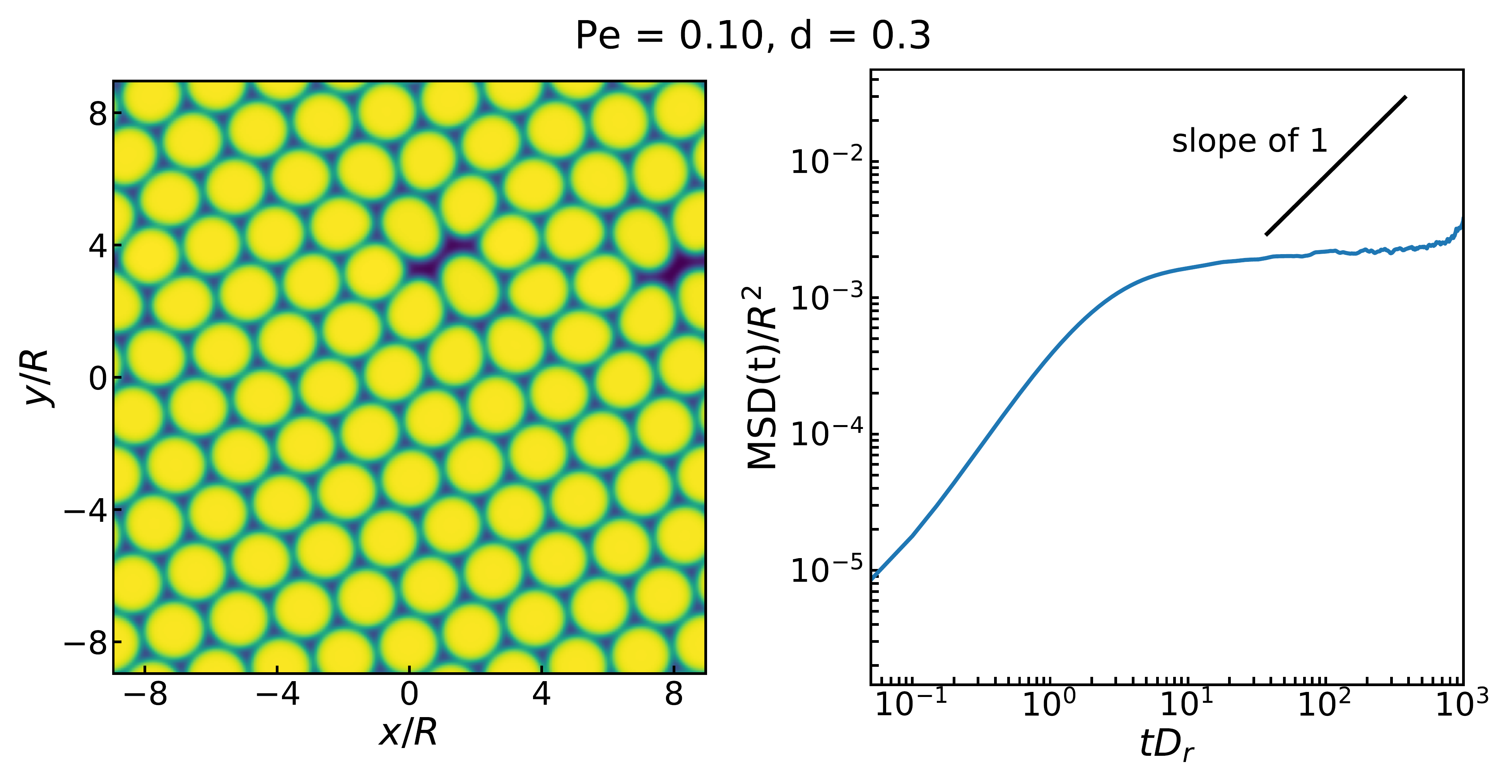}
    \includegraphics[width=0.49\linewidth]{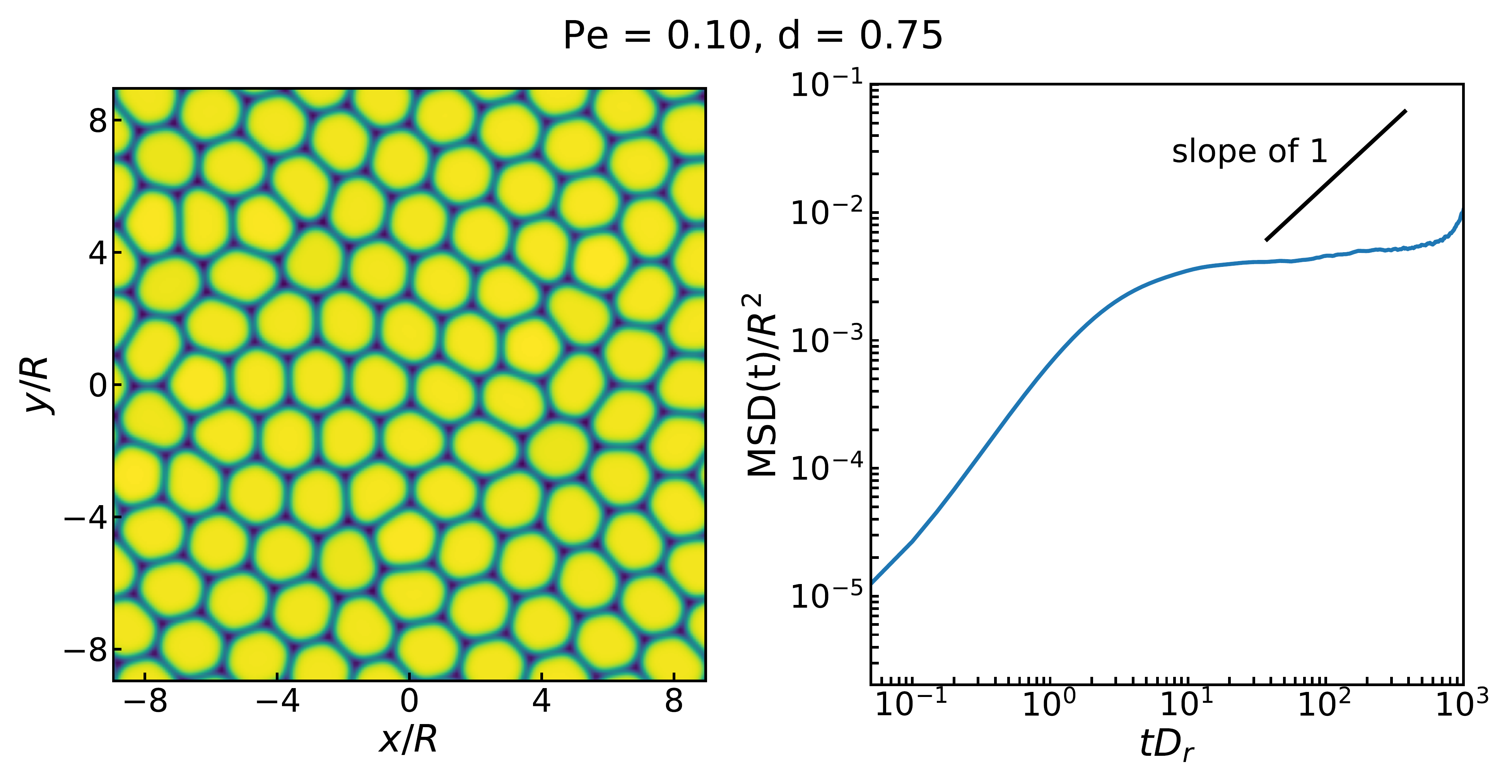}
    \includegraphics[width=0.49\linewidth]{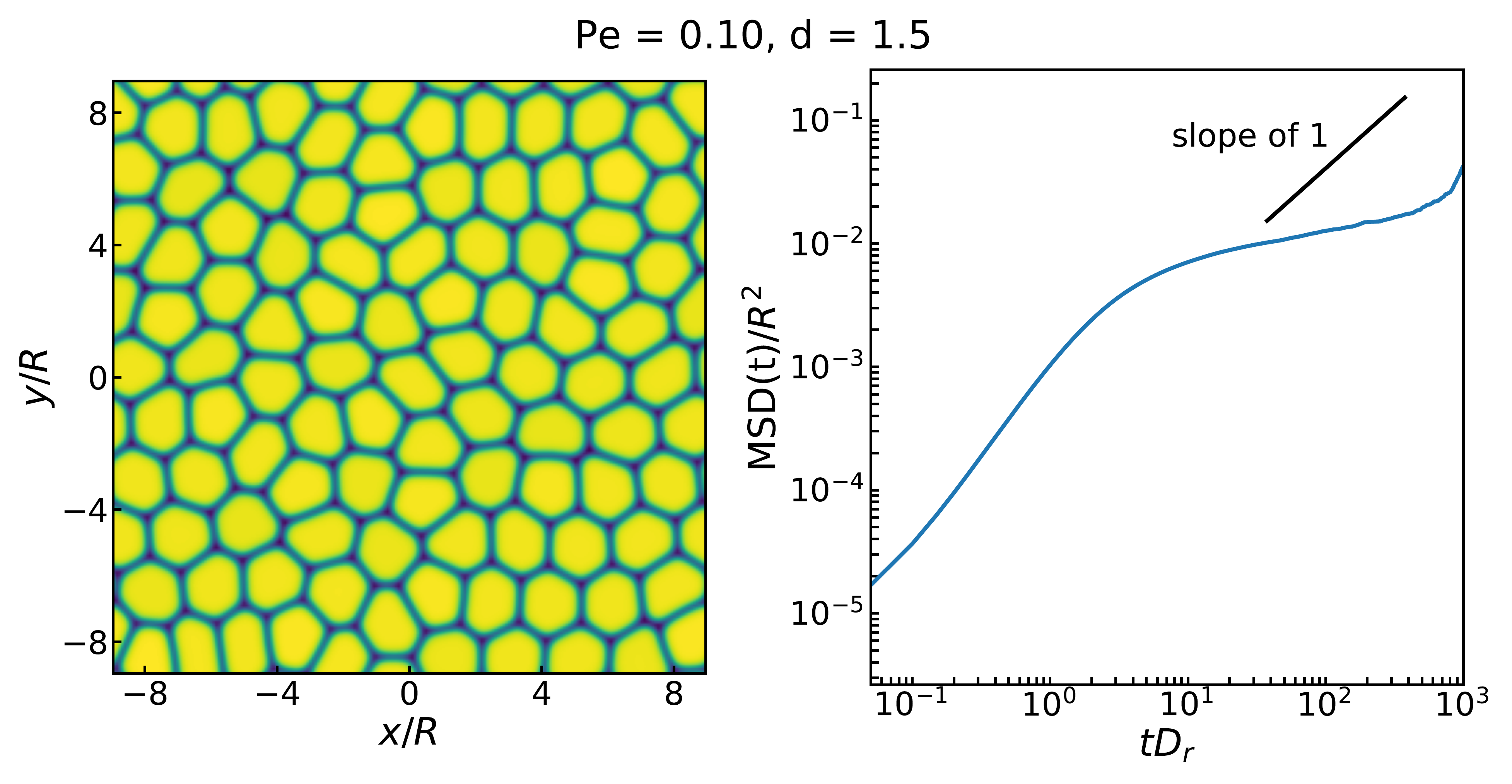}
    \includegraphics[width=0.49\linewidth]{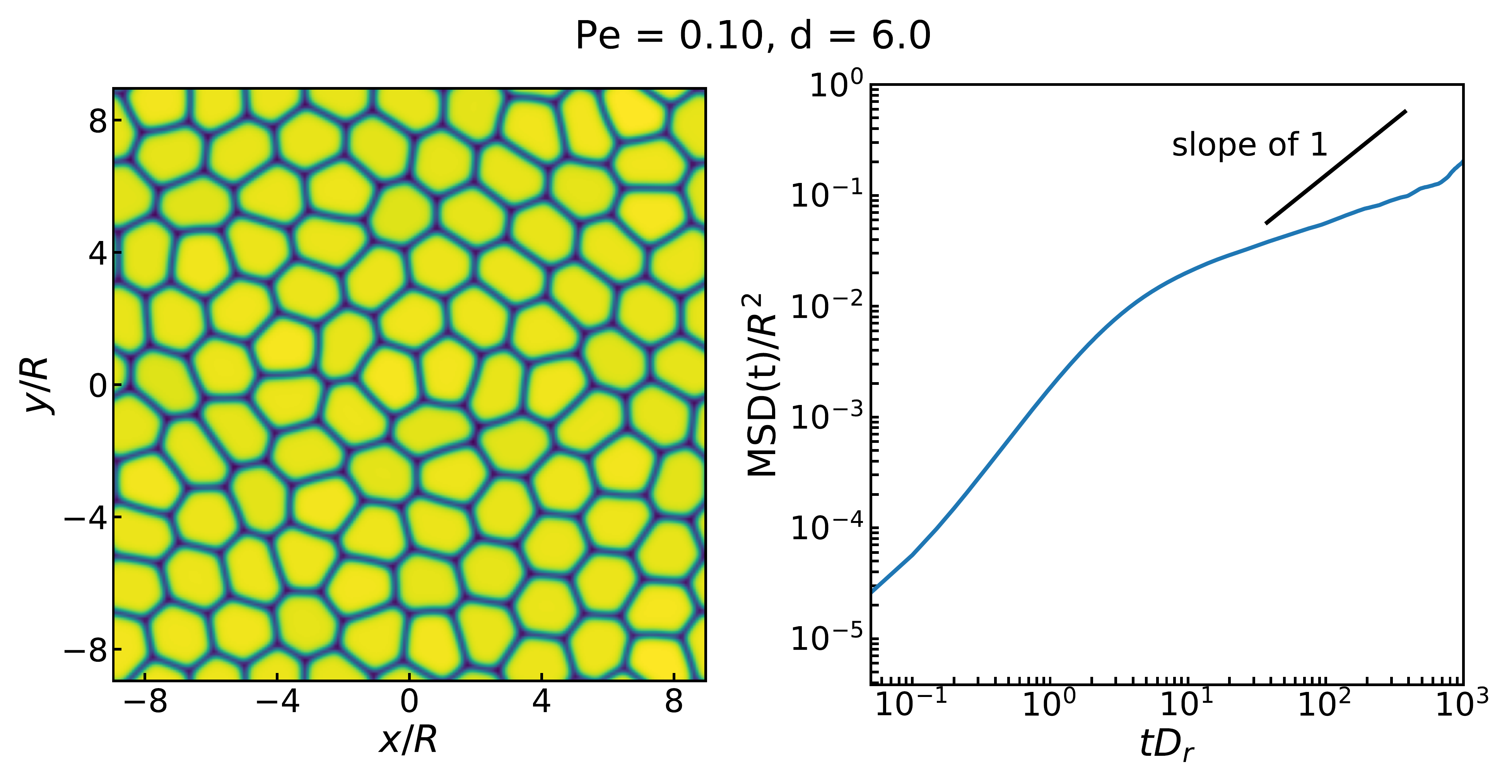}
    \caption{Snapshots and mean squared displacements (MSDs) \mc{for  systems with fixed $\text{Pe} = 0.1$ and $d$ varied from $0.3$ to $6$, as considered in \fig\ref{fig:fc_v0_and_d}(b). Across all cases the slope of the MSDs remains far below $1$, suggesting that the monolayer remains solid-like.}} 
    \label{fig:msds_d}
\end{figure*}

In \mc{\fig\ref{fig:msds_v0} and \fig\ref{fig:msds_d}}, we report the mean squared displacements of the systems we consider in the main paper, classifying them as glassy for subdiffusive MSDs and liquid for diffusive ones.
When determining the mean squared displacement, we consider the system without a probe.

\section*{The critical force}
\begin{figure*}[!th]
\includegraphics[width=0.75\textwidth]{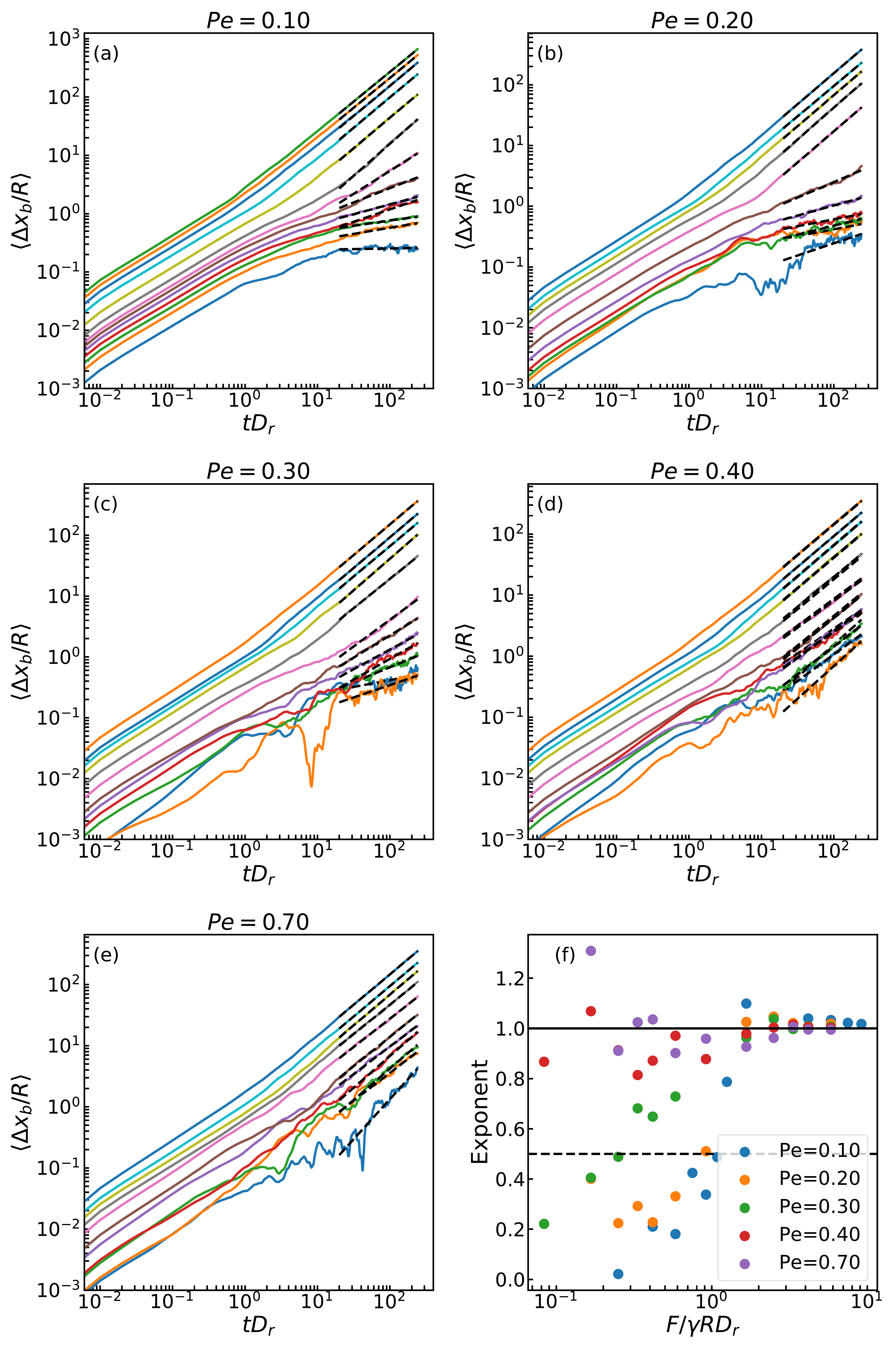}
\caption{(a)--(e) show the displacement $\Delta x_b$ of the probe in the force direction, averaged over different initial configurations, for various values of the tissue self-propulsion at fixed tissue deformability $d=3.0$. The dashed lines are linear fits to extract the long time exponent. (f) shows the best fit exponents for the long time data, excluding the points at the lowest forces where such a fit is not reliable. The solid and dashed lines show reference values of 1.0 and 0.75 respectively.}
\label{fig:xb_Pe}
\end{figure*}

\begin{figure*}[!th]
\includegraphics[width=0.75\textwidth]{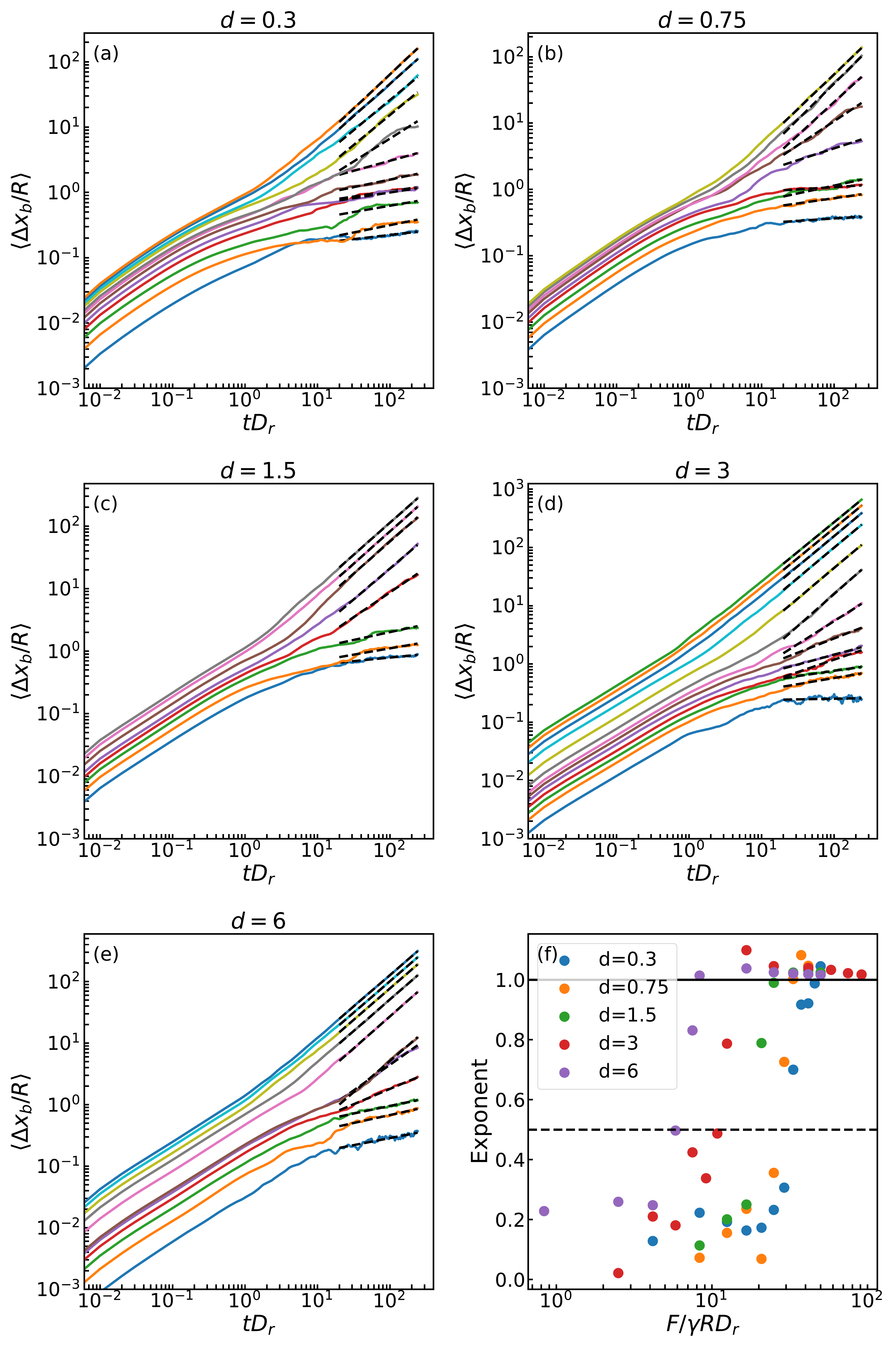}
\caption{As in \fig \ref{fig:xb_Pe}, but for various values of the tissue deformability at fixed tissue self-propulsion $\text{Pe} = 0.10$.}
\label{fig:xb_d}
\end{figure*}

In \mc{\fig\ref{fig:xb_Pe} and \fig\ref{fig:xb_d}}, we show the change in the probe's position ($\Delta x_b(t) = x_b(t) - x_b(0)$) over time for various forces, averaged over the runs. The exponent of the long time portion of the curve characterizes the dynamics of the probe. An exponent is 0 for a stationary probe, 1 for a probe moving steadily, and values in between those for stick-slip motion. In colloidal systems, this exponent has been related to the critical force, although simulations tend not to level off as flatly as related mode-coupling calculations \cite{Gruber2016,Gruber2020}.

\begin{figure*}[!th]
\includegraphics[width=0.75\textwidth]{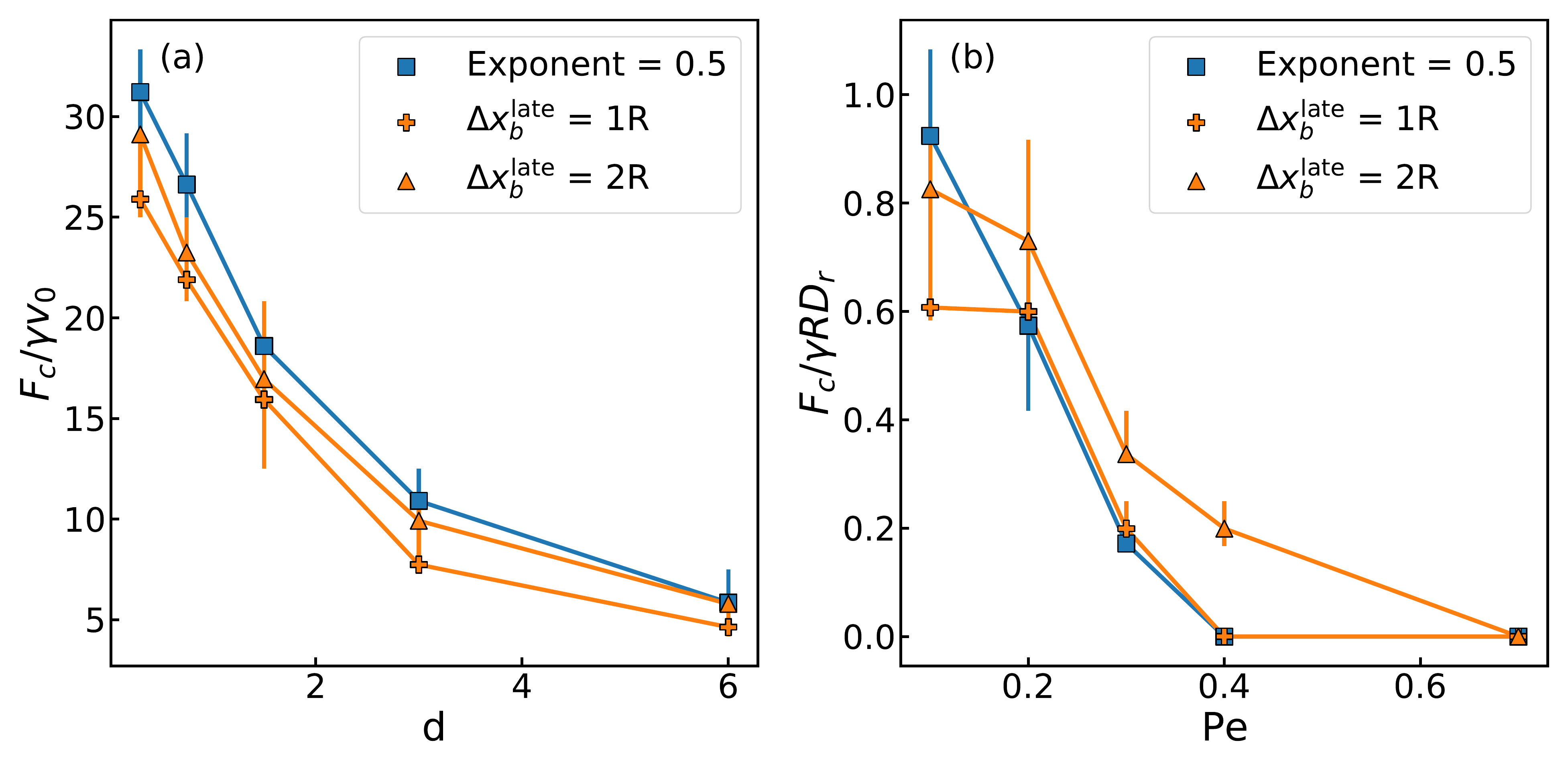}
\caption{The threshold force as determined by both the long time exponent of the position-time curve and the long time displacement of the bead. In (a) the deformability is varied at a fixed $\text{Pe} = 0.10$. In (b) the self-propulsion is varied at a fixed $d=3.0$. Note the difference in scale between the two: $v_0$ = 0.10$RD_r$.}
\label{fig:fc_sup}
\end{figure*}

Therefore, in \fig\ref{fig:fc_sup}, we characterize two approaches to defining a critical force $F_c$, based on the position of the probe as a function of time. One is to choose a cutoff based on the exponent of $\Delta x_b(t)$ at long times and the second is to choose a cutoff based on the total long-time change in position. We show a comparison between the two methods in \fig\ref{fig:fc_sup}. In the main paper, we report results based the cutoff of steady state displacement of at least $1R$. As can be seen in \fig\ref{fig:fc_sup}(a), the trends with deformability are the same for both methods of characterizing $F_c$ and do not vary strongly with the exact choice of the threshold.

\section*{Radial compression at low deformability}
\begin{figure*}[!ht]
    \centering
    \includegraphics[width=0.75\linewidth]{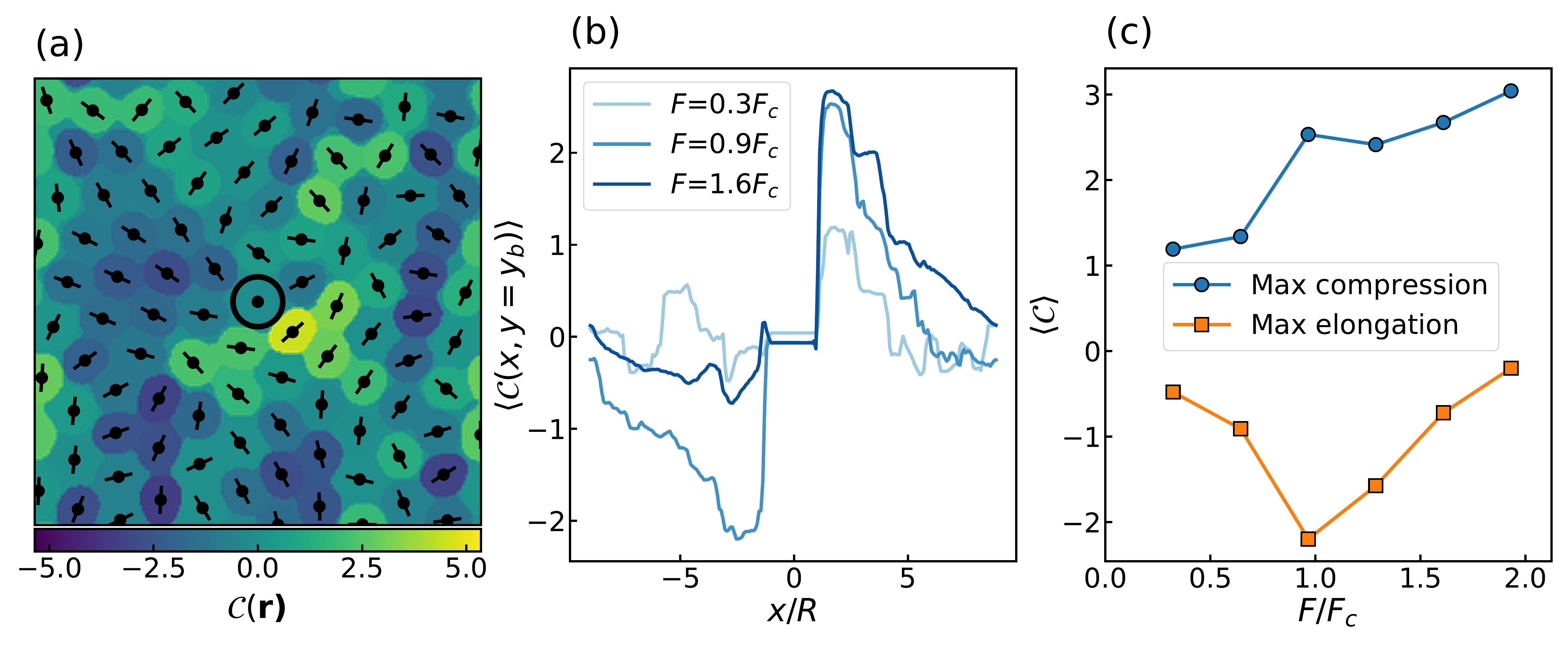}
    \caption{Similar to \fig~\ref{fig:def_f}, but for a low deformability $d = 0.3$. Note that the scale is different, the magnitude of $\mathcal{C}$ tends to be lower because the particles do not deform as much. (a) A snapshot of the radial compression field for $F=0.84F_c$. (b) Profile of the compression field along the center of the probe in the $x$ direction, i.e., $\langle\mathcal{C}(x,y=y_b)\rangle$. Darker shades of blue correspond to greater forces. (c) The minima and maxima of $\langle\mathcal{C}(x,y=y_b)\rangle$ as a function of the applied force. }
    \label{fig:def_f2}
\end{figure*}

In \mcN{\fig\ref{fig:def_f2}} we show the radial compression $\mathcal{C}$ for a low deformability $d=0.3$ glassy state. The magnitude of the deformations is much smaller than for the $d=3$ system in the \fig\ref{fig:def_f}. The elongation is especially reduced because low deformability cells tend not to elongate as much to fill empty spaces, which is also reflected in the gaps that occur in this system. However, the maximum compression in front of the probe monotonically increases with force in a similar fashion as in the $d=3$ system.

\begin{figure*}[!th]
\includegraphics[width=0.85\textwidth]{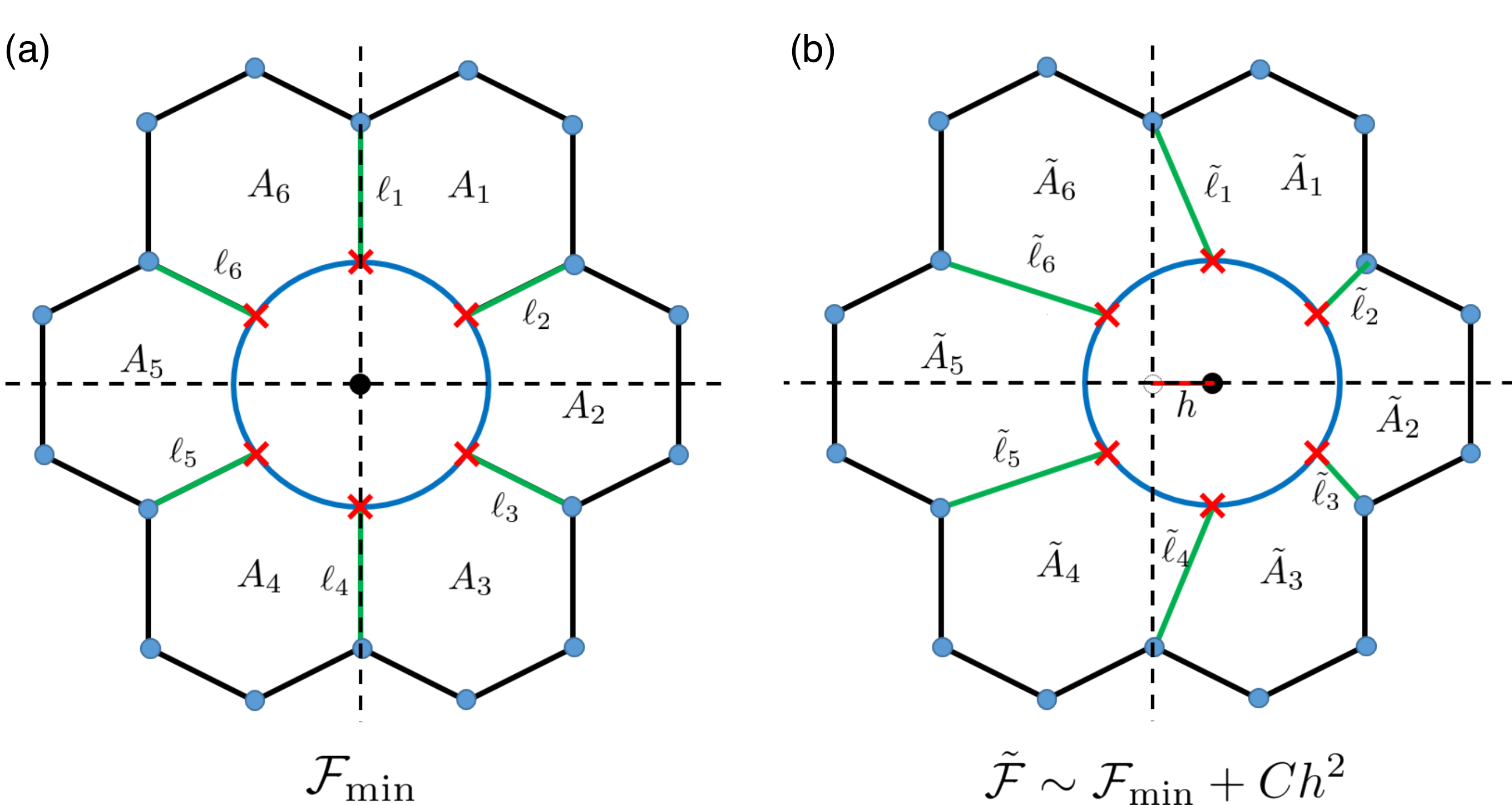}
\caption{\bl{\mcm{Configuration of a probe particle caged by surrounding cells.}  The total free energy of the system can be approximated in terms of the \mcm{deviations of the} perimeters and areas of the surrounding cells \mcm{from their rest values}. (a) At rest, the probe sits in a configuration of minimum free energy. (b) Moving the probe \mc{by} a small distance $h$ deforms the surrounding cells, leading to changes in their perimeters and areas, and consequently, to a higher value of the energy. Since this displacement distorts a minimum of the free energy, the associated change in energy must be quadratic in $h$.}}
\label{fig:effK}
\end{figure*}

\bl{

\section*{Effective Spring Constant}

In this section, we estimate the effective spring constant $k_{\text{eff}}$ felt by a caged \mcm{cell}. For simplicity, let us consider the setup \mc{shown in} \fig\ref{fig:effK}(a), \mcm{where the probe particle is} surrounded by six cells in a relaxed state, so the system is in a local \mcm{minimum} of the free energy $\mathcal{F} = \mathcal{F}_\text{min}$. 

We \mcm{now examine the deformation energy obtained when the probe particle} is translated a distance $h$, \mc{as depicted in} \fig\ref{fig:effK}(b), \mcm{resulting in an}
increase in free energy to $\mathcal{F} = \tilde{\mathcal{F}} > \mathcal{F}_\text{min}$.
The free energy has three contributions:
\begin{equation}
    \mathcal{F} = \mathcal{F}_\text{Overlap} +  \mathcal{F}_\text{Tension} + \mathcal{F}_\text{Compression}.
\end{equation}
Assuming that \mcm{cell pairs overlap in a region} of the order of the interfacial thickness $\xi$, the overlap contribution is proportional to the total length $L_T$ of their shared sides,
\begin{equation}
    \mathcal{F}_\text{Overlap} = \xi \epsilon \sum_n \ell_n = \xi \epsilon L_T,
\end{equation}
where $\epsilon$ is the strength of repulsion and $ \ell_n$ is the length of  shared sides of cell $n$.  Similarly, as cells have surface tension $\sigma$, each of these sides also contributes with a tension term to the free energy
\begin{equation}
    \mathcal{F}_\text{Tension} = 2\sigma \sum_n \ell_n = 2\sigma L_T,
\end{equation}
where the factor of $2$ comes from the fact that each side is shared by two cells. Finally, cells have a preferred area \mcm{$A_0=\pi R^2$}. Deviations from this value cost an energy  
\begin{equation}
    \mathcal{F}_\text{Compression} = \frac{\lambda}{A_0^2}\sum_n (A_n-A_0)^2 = \frac{\lambda}{A_0^2} \sum_n A^2_n + \text{const.},
\end{equation}
where $\lambda$ is the strength of the soft constraint on the area of each cell, and we have used the fact that for a confluent tissue $\sum_n A_n$ is constant, and remains so even when the probe particle moves. \mcm{The total free energy can then be written as}
\begin{equation}
\label{eq:all_contrb}
    \mathcal{F} = (\xi\epsilon+2\sigma) L_T + \frac{\lambda}{A_0^2} A^2,
\end{equation}
where we have defined $A^2\equiv\sum_nA_n^2$. \mcm{When} we move the probe \mcm{away from} its equilibrium position, the free energy increases by
\begin{equation}
    \delta\mathcal{F} = \tilde{\mathcal{F}}-\mathcal{F}_\text{min} =  (\xi\epsilon+2\sigma) \delta L_T + \frac{\lambda}{A_0^2} \delta A^2,
\end{equation}
where $\delta L_T = \tilde{L}_T-L_{T,\text{min}}$ and  $\delta A^2 = \tilde{A}^2-A^2_\text{min}$.

As the original configuration was a local minimum of the free energy, \mcm{a small displacement $h$ of the probe particle from} this configuration must lead to a  correction quadratic in $h$, i.e., $\delta \mathcal{F} \sim h^2$. \mcm{Since} the parameters $\xi$, $\epsilon$, $\sigma$ and $\lambda$ can be varied independently, \eq\eqref{eq:all_contrb} implies that we must have $\delta L_T = c_1 h^2/(2 R)$ and $\delta A^2 = c_2 \pi^2 R^2 h^2/2$, where $c_1$ and $c_2$ are numerical coefficients that depend on the details of the initial configuration and, possibly, on the orientation of the displacement. \mcm{We can then write}
\begin{equation}
    \delta\mathcal{F} = \frac{1}{2}\left(c_1\frac{\xi\epsilon+2\sigma}{R} + c_2\frac{\lambda}{R^2}\right) h^2,
\end{equation}
which leads to the following \mcm{restoring} force on the \mcm{probe particle}
\begin{equation}
    F = -\partial_h \delta \mathcal{F} = -\left(c_1\frac{\xi\epsilon+2\sigma}{R} + c_2\frac{\lambda}{R^2}\right) h,
\end{equation}
from which we extract the effective spring constant
\begin{equation}
    k_\text{eff} = c_1\frac{\xi\epsilon+2\sigma}{R} + c_2\frac{\lambda}{R^2}.
\end{equation}
In our simulations, we keep $\epsilon$ and $\xi$ constant. Therefore we can rewrite $k_\text{eff}$ as
\begin{equation}
    k_\text{eff} = \frac{\xi\epsilon}{R}\left(c_1\left(1+\frac{2}{d}\right) + c_2\chi\right),
        \label{eq:keff}
\end{equation}
where $d = \epsilon\, \xi/\sigma$ is the deformability and $\chi = \lambda/(\epsilon\xi R)$ is the compressibility.  The result given in \eq\eqref{eq:keff} shows that the tissue becomes stiffer by either reducing deformability (\mcm{controlled by cell surface tension and} associated \mcm{with} changes in cell perimeter) or by increasing  compressibility (associated \mcm{with} changes in size or area). Since $\epsilon$ sets the scale of any interaction, its presence in the prefactor of \eq\eqref{eq:keff} is justified. 
}

\onecolumngrid
\newpage
\section*{Supplemental Figures}

\begin{figure*}[!ht]
    \centering
    \includegraphics[width=0.6\linewidth]{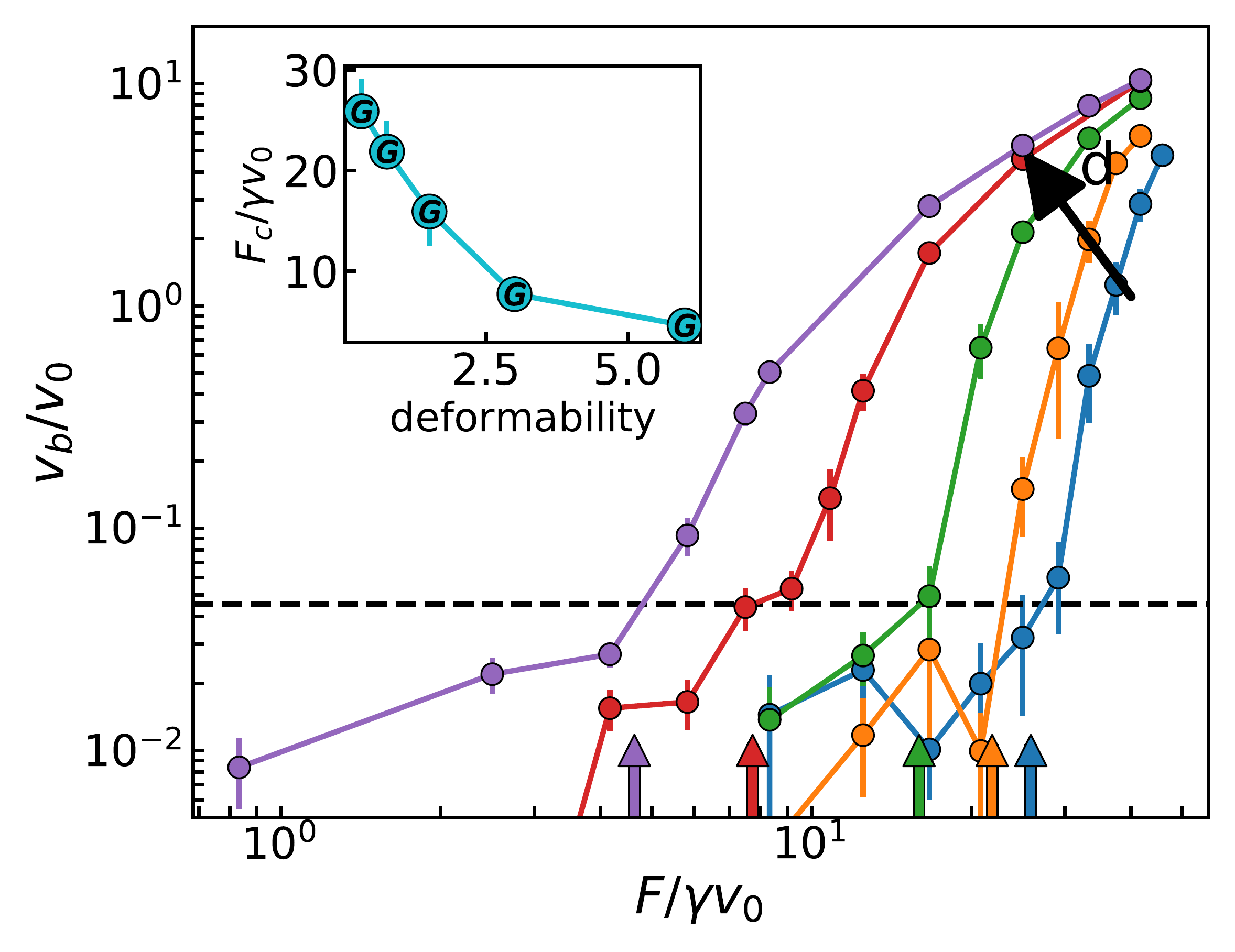}
    \caption{\mcm{The data of \fig~\ref{fig:fc_v0_and_d}(b) are plotted here on log-log scale to highlight the sharpness of the transition at low deformability.} The black dashed line indicates the velocity cutoff corresponding to $\Delta x_b^{\textrm{late}} = R$.}
    \label{fig:fc_d2}
\end{figure*}


\begin{figure*}[!ht]
    \centering
    \includegraphics[width=0.6\linewidth]{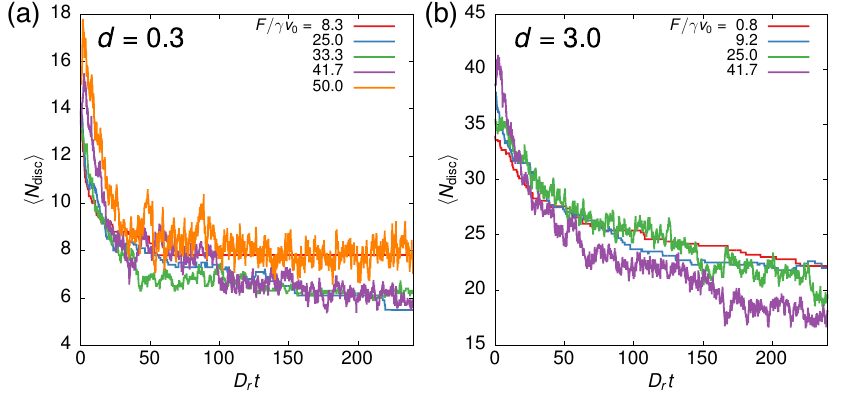}
    \caption{The average number of disclinations $\langle N_{\text{disc}}\rangle$ in the system over the simulation period for (a) $d = 0.3$ and (b) $d = 3.0$ at various applied forces. \mc{Note that $\langle N_{\text{disc}}\rangle$ exhibits larger fluctuations at higher forces.}}
    \label{fig:defects_time}
\end{figure*}

\end{document}